\documentclass[aps,prb,superscriptaddress,twocolumn]{revtex4-2}
\usepackage{amsfonts, amssymb, amsmath,amsthm,graphicx}
\usepackage{mathrsfs}
\usepackage{subfigure}
\usepackage{multirow}
\usepackage{dcolumn}
\usepackage{bm}
\usepackage{color}
\usepackage[usenames,dvipsnames]{xcolor}
\usepackage[all]{xy}
\usepackage{enumitem}

\renewcommand{\bf}[1]{\textnormal{\textbf{#1}}}
\newcommand{\BZ}{\textnormal{\text{BZ}}}

\newcommand{\directint}{\int^{\oplus}}

\newcommand{\thetachar}[2]{\vartheta \left[\begin{array}{c}
#1\\
#2
\end{array}\right]}

\newtheorem{proposition}{Proposition}
\newtheorem{theorem}{Theorem}

\newcommand{\ket}[1]{| #1 \rangle}
\newcommand{\bra}[1]{\langle #1|}

\usepackage[colorlinks=true,allcolors=blue]{hyperref}

\begin{document}

\author{Bruno Mera}

\affiliation{Advanced Institute for Materials Research (WPI-AIMR), Tohoku University, Sendai 980-8577, Japan}

\affiliation{Instituto de Telecomunica\c{c}\~oes and Departmento de Matem\'{a}tica, Instituto Superior T\'ecnico, Universidade de Lisboa, Avenida Rovisco Pais 1, 1049-001 Lisboa, Portugal}

\author{Tomoki Ozawa}

\affiliation{Advanced Institute for Materials Research (WPI-AIMR), Tohoku University, Sendai 980-8577, Japan}

\title{Uniqueness of Landau levels and their analogs with higher Chern numbers}
\date{\today}
\begin{abstract}
Landau levels are the eigenstates of a charged particle in two dimensions under a magnetic field, and are at the heart of the integer and fractional quantum Hall effects, which are two prototypical phenomena showing topological features. Following recent discoveries of fractional quantum Hall phases in van der Waals materials, there is a rapid progress in understanding of the precise condition under which the fractional quantum Hall phases can be stabilized. It is now understood that the key to obtaining the fractional quantum Hall phases is the energy band whose eigenstates are holomorphic functions in both real and momentum space coordinates. Landau levels are indeed examples of such energy bands with an additional special property of having flat geometrical features. In this paper, we prove that, in fact, the only energy eigenstates having holomorphic wave functions with a flat geometry are the Landau levels and their higher Chern number analogs. Since it has been known that any holomorphic eigenstates can be constructed from the ones with a flat geometry such as the Landau levels, our uniqueness proof of the Landau levels allows one to construct any possible holomorphic eigenstate with which the fractional quantum Hall phases can be stabilized. 
\end{abstract}
\maketitle 
\section*{Introduction}
There is an increasing interest in topological flat bands. Here, the adjective \textit{flat} may refer not only to the energy dispersion but also to the geometrical properties in momentum space such as the Berry curvature and the quantum metric. The lowest Landau level is an example of such an energetically and geometrically flat band. Additionally, the lowest Landau level has a special property that the wave function can be taken as a holomorphic function both in real and momentum space~\cite{YoshiokaBook,wang:cano:millis:liu:yang:21}. It has been known that when bands fulfill the holomorphicity condition and the Berry curvature is flat, their projected density operators obey the Girvin-MacDonald-Platzman (GMP) algebra~\cite{girvin:macdonald:platzman:86}, which implies that a fractional topological phase can be stabilized under short-range interactions~\cite{roy:14}.
Recently, it has been found that the holomorphicity is the key to obtaining the fractional topological phases, and the strict flatness of the Berry curvature is not necessary~\cite{wang:klevstov:liu:22}. It is known that any holomorphic state can be constructed by modifying holomorphic wave functions with uniform Berry curvature. Twisted bilayer graphene in the chiral limit~\cite{ bistritzer:macdonald:11, cao:fatemi:fang:18, cao:fatemi:demir:18,andrei:macdonald:2020,ledwith:tarnopolsky:khalaf:vishwanath:20,wang:liu:22,arbeitman:chew:bernevig:22} provides an example of such a holomorphic state. With the recent discovery of fractional quantum Hall states in bilayer graphene~\cite{Spanton:2018Science,Xie:2021Nature} and fractional quantum anomalous Hall states in twisted moiré lattices~\cite{FQAH_Xu1,FQAH_Xu2,FQAH_Jie,FQAH_Li,FQAH_Ju}, it is of urgent interest to identify the class of wave functions where the fractional topological phases can be stabilized.

In this paper, we show that the lowest Landau level is not just an example of geometrically flat bands fulfilling the holomorphicity condition, but rather they are the \textit{only} possibility of such an isotropic flat band in two dimensions with unit Chern number. We also show that geometrically flat bands with higher Chern numbers, such as those constructed in~\cite{wu:regnault:bernevig:13}, are also uniquely determined once one fixes the Chern number and one parameter governing the anisotropy called the \emph{modular parameter}. Finally we provide explicit expressions of those wave functions. 
On the one hand, our result places a strict constraint on the types of wave functions one can consider in exploring bands fulfilling the GMP algebra. On the other hand, our explicit expressions of the desired wave functions, which are exhaustive due to the uniqueness, can provide a solid basis to further study fractional topological physics in such bands.

This paper is structured as follows. In Sec.~\ref{sec: results} we present our results. We begin by recalling and discussing the recently introduced concept of K\"ahler bands in Sec.~\ref{subsec: Kaehler bands}, then we present our main theorem in Sec.~\ref{subsec: main theorem}, and in Sec.~\ref{subsec: proof} we present its proof. In Sec.~\ref{sec: discussion} we discuss our results. In Sec.~\ref{sec: technical results} we present some technical results used in the proof of the main theorem.
\section{Results}
\label{sec: results}
\subsection{K\"{a}hler bands}
\label{subsec: Kaehler bands}
Before stating our main theorem and presenting its proof, we first introduce the basic terminology of Bloch bands and the concept of K\"ahler band which will be used in the discussion to follow. We focus on two spatial dimensions. 

Energy bands of a particle in a periodic potential are characterized, via Bloch's theorem, by the quasimomentum $\mathbf{k}$, a parameter which is taken from a two-torus known as the Brillouin zone $\BZ^2$. For a given $\mathbf{k}$, an eigenstate for a given band is described by a Bloch wave function $|u_\mathbf{k}\rangle$ which takes values in a suitable Hilbert space $\mathcal{H}$. Since multiplying $|u_\mathbf{k}\rangle$ by a nonzero complex number defines the same quantum state, a quantum state is specified by a one-dimensional vector subspace of $\mathcal{H}$ i.e., a point in a complex projective space. A Bloch band thus defines a map from the Brillouin zone to a complex projective space. 

In Refs.~\cite{ozawa:mera:21,mera:ozawa:21,mera:ozawa:21:engineering}, we have introduced the concept of a \emph{(anti)-K\"{a}hler band}, a Bloch band for which the associated map from the Brillouin zone to the space of quantum states, regarded appropriately as a map between complex manifolds, is a  (anti)-holomorphic map with nonvanishing derivative [mathematically, the map is a (anti)-holomorphic immersion]. For (anti)-K{\"a}hler bands, the \emph{quantum metric} $g$ and \emph{Berry curvature} $F$ are connected by a \emph{complex structure} $J$, a tensor which squares to $-1$, giving $\BZ^2$ the structure of a \emph{(anti)-K\"{a}hler manifold}.
In the periodic coordinates $\bf{k}=(k_x,k_y)$ of the $\BZ^2$~\footnote{we use a convention for which the irreducible representations (irreps) of the real space lattice $\mathbb{Z}^2$ are written as $e^{2\pi i\bf{k}\cdot\bf{\bf{R}}}$, with $\bf{R}\in\mathbb{Z}^2$, so that $\bf{k}$ and $\bf{k} + \bf{G}$, with $\bf{G}\in \mathbb{Z}^2$, determine the same irrep and hence the same Bloch quasimomentum in $\BZ^2$.}, denoting the two-by-two matrices representing $g$, $F$ and $J$, respectively, by the same letters $g(\bf{k})=[g_{ij}]_{1\leq i,j\leq 2}$, $F(\bf{k})=[F_{ij}]_{1\leq i,j\leq 2}$ and $J(\bf{k})=[J_{ij}]_{1\leq i,j\leq 2}$, the condition that a K\"{a}hler band must satisfy is equivalent to the matrix equation
\begin{align}
g(\bf{k})=-\frac{i}{2}F(\bf{k}) J(\bf{k}), \label{eq:equality1}
\end{align}
which taking determinants gives the quantum metric-Berry curvature relation
\begin{align}
\sqrt{\det(g(\bf{k}))}=\frac{|F_{12}(\bf{k})|}{2}. \label{eq:equality2}
\end{align}
(Note that we use the convention that the Berry curvature $F$ is purely imaginary.)
In Ref.~\cite{ozawa:mera:21}, the lowest Landau level appears as a special case of a K\"{a}hler band with $\mathcal{C}=1$ for which the quantum geometry of the Brillouin zone is independent of $\mathbf{k}$, i.e., it is invariant under the full translation group $\mathbb{R}^2$ of the Brillouin zone---we refer to these Bloch bands as \emph{geometrically flat K\"{a}hler bands}~\footnote{We note that concepts closely related to K{\"a}hler bands have been proposed by several authors, and we would like to clarify their relations here. \emph{Ideal flatbands}~\cite{wang:cano:millis:liu:yang:21,wang:klevstov:liu:22} refer to energetically flat K{\"a}hler bands with a constant complex structure $J$. \emph{Vortexable bands} refer to bands where one can add vortices without leaving the eigenspace of bands; in the presence of a lattice translation symmetry, vortexable bands are equivalent to ideal flatbands~\cite{ledwith:vishwanath:parker:2022}.}.
\subsection{Main theorem}
\label{subsec: main theorem}
The main result of this paper is the following theorem:
\begin{theorem} [Uniqueness of geometrically flat K\"{a}hler bands]
Geometrically flat (anti-)K\"ahler bands i.e., bands in which the equality Eq.~(\ref{eq:equality1}) holds and $g$ is independent of $\bf{k}$, are \emph{unique} up to a gauge choice, given the Chern number $\mathcal{C}\in\mathbb{Z}_{\neq 0}$ and the constant complex structure $J$.
\label{th: uniqueness of flat Kaehler bands}
\end{theorem}
Some remarks are in order. The cases $\mathcal{C} > 0$ corresponds to K\"ahler bands, while $\mathcal{C} < 0$ corresponds to anti-K\"ahler bands. Below we assume $\mathcal{C} > 0$, but the uniqueness proof holds for $\mathcal{C} < 0$ \textit{mutatis mutandis}; in particular, ones obtains the wave function for $\mathcal{C} < 0$ by taking the complex conjugate of the wave function for $\mathcal{C} > 0$. 

The assumption that $g$ is constant, together with Eq.~(\ref{eq:equality2}), implies that $F$ is also constant in momentum space. Besides, from Eq.~(\ref{eq:equality1}), one sees that $J$ is also constant. Such a K\"ahler band is thus \textit{translation-invariant} in momentum space.
Translation-invariant complex structures $J$ in the Brillouin zone $\BZ^2$ can be conveniently parametrized in terms of a modular parameter $\tau \in\mathbb{H}$ as $J = \dfrac{1}{\mathrm{Im}(\tau)}\begin{pmatrix} -\mathrm{Re}(\tau) & -|\tau|^2 \\ 1 & \mathrm{Re}(\tau)\end{pmatrix}$, where $\mathbb{H}=\{\tau\in\mathbb{C} : \mathrm{Im}(\tau) >0\}$ is the upper half of the complex plane~\cite{mera:ozawa:21:engineering}.
We then parametrize momentum space by the complex coordinate $z_{\tau}=k_x +\tau k_y$. Note that the Brillouin zone equipped with this complex coordinate becomes a complex torus $\mathbb{C}/\Lambda_{\tau}$, where $\Lambda_{\tau}=\mathbb{Z}+\tau\mathbb{Z}$.

We will see that one cannot obtain a geometrically flat K\"{a}hler band with a finite number of total bands---a result which was already alternatively proved in Refs.~\cite{mera:ozawa:21:engineering, varjas:abouelkomsan:yang:bergholtz:22}.

The outline of the proof of the Theorem is the following. We first consider K\"ahler bands with constant $J$ and show that Bloch wave functions must be written as a linear combination of theta functions with characteristics. We then assume that $g$, and hence $F$, are constant and, via the Stone-von Neumann theorem, show that only one possible combination of theta functions is allowed.
\subsection{Proof}
\label{subsec: proof}
{\it K\"{a}hler bands for translation-invariant $J$.---} We describe the Bloch wave function~\footnote{Here the Bloch wave function refers to the unit-cell periodic (up to a suitable phase when under a magnetic field) part of the eigenstate with a given quasimomentum.} in two spatial dimensions by a collection of nonvanishing vectors $\ket{u_{\bf{k}}}\in \mathcal{H}$ smoothly parametrized by $\bf{k}\in\mathbb{R}^2$ where $\mathbb{R}^2$ is the universal cover of $\BZ^2$, and $\mathcal{H}$ is a \emph{fixed} Hilbert space which can be finite or infinite dimensional. We need $\ket{u_{\bf{k}}}$ and $\ket{u_{\bf{k}+\bf{G}}}$ for arbitrary $\bf{G}$ in the reciprocal lattice to define the same quantum state. If the momentum space Hamiltonian $H_\mathbf{k}$ obeys the periodicity $H_{\mathbf{k}+\mathbf{G}} = H_{\mathbf{k}}$, $\ket{u_{\bf{k}}}$ and $\ket{u_{\bf{k}+\bf{G}}}$ differ, in general, by multiplication by a non-vanishing complex number $e_{\bf{G}}(\bf{k})$:
\begin{align}
\ket{u_{\bf{k}+\bf{G}}}=\ket{u_{\bf{k}}} e_{\bf{G}}(\bf{k}).
\label{eq: global multivalued frame field for Bloch bundle}
\end{align}
By associativity of the sum (see Eq.~\eqref{eq: cocycle condition for multipliers and unitary rep of reciprocal lattice}), one can show that the functions $e_{\bf{G}}(\bf{k})$ satisfy a cocycle condition
\begin{align}
e_{\bf{G}_1+\bf{G}_2}(\bf{k})=e_{\bf{G}_1}(\bf{k}+\bf{G}_2)e_{\bf{G}_2}(\bf{k}).
\end{align}
The family of functions then defines what is called a \emph{system of multipliers} for a line bundle $L\to\BZ^2$ (see, for example, Refs.~\cite{mumford:74,beauville:13}). There is a gauge degree of freedom in defining the Bloch wave function $\ket{u_{\bf{k}}}$ because a state in quantum mechanics is a one-dimensional subspace of the Hilbert space. Below, we will indeed frequently use non-normalized vectors because it is convenient in the holomorphic setting. Concretely, we are free to multiply $\ket{u_{\bf{k}}}$ by $g(\bf{k})\in \mathrm{GL}(1;\mathbb{C})=\mathbb{C}^*$ depending smoothly on $\bf{k}\in\mathbb{R}^2$. This changes the multipliers to $\frac{g(\bf{k}+\bf{G})}{g(\bf{k})}e_{\bf{G}}(\bf{k})$, which does not change the isomorphism class of $L$~\cite{beauville:13}. We assume that this line bundle has Chern number $\mathcal{C}$, which is nothing but the Chern number of the band under consideration.

More generally, the momentum-space Hamiltonian can transform as $H_{\mathbf{k}+\mathbf{G}} = V_{\mathbf{G}}H_{\mathbf{k}}V_{\mathbf{G}}^{-1}$, where, as a consequence of Wigner's theorem~\cite{freed:12}, $V_{\mathbf{G}}$ is either a unitary or anti-unitary transformation corresponding to a projective representation of the symmetry group consisting of the reciprocal lattice $\mathbb{Z}^2$. Physically, $V_{\mathbf{G}}$ is determined by the spatial structure within a unit cell, which can affect the resulting quantum geometry~\cite{Huhtinen:2022}. Therefore, there can be a collection of unitary or anti-unitary transformations $V_{\bf{G}}$ of $\mathcal{H}$, $\bf{G}\in\mathbb{Z}^2$, such that
\begin{align}
\ket{u_{\bf{k}+\bf{G}}}=\left(V_{\bf{G}}\ket{u_{\bf{k}}}\right)e_{\bf{G}}(\bf{k}),
\label{eq:quasiperiodic}
\end{align}
and such that $V_{\bf{G}_1+\bf{G}_2}$ equals to $V_{\bf{G}_1}V_{\bf{G}_2}$ up to a phase factor which depends on $\bf{G}_1$ and $\bf{G}_2$. These matrices can not depend on $\bf{k}$ otherwise the Berry curvature and quantum metric, which are tensors in the Brillouin zone, would not be periodic in $\bf{k}$ with respect to reciprocal lattice translations. Since an anti-unitary $V_{\bf{G}}$ would necessarily change the sign of the Berry curvature implying the existence of zero of the Berry curvature, contradicting with the assumption~\footnote{The property that the K{\"a}hler band is an immersion to the complex projective space implies that the quantum metric must be non-degenerate over the entire Brillouin zone, and hence $\sqrt{\det (g)} \neq 0$ and thus $|F_{12}| \neq 0$.}, the projective representation must be unitary. Additionally, we note that, by associativity of the sum,
\begin{align}
\label{eq: cocycle condition for multipliers and unitary rep of reciprocal lattice}
&\ket{u_{\bf{k}+\left(\bf{G}_1+\bf{G}_2\right)}}=\left(V_{\bf{G}_1+\bf{G}_2}\ket{u_{\bf{k}}}\right)e_{\bf{G}_1+\bf{G}_2}(\bf{k}) \\
=&\ket{u_{\left(\bf{k}+\bf{G}_2\right) +\bf{G}_1}} =\left(V_{\bf{G}_1}V_{\bf{G}_2}\ket{u_{\bf{k}}}\right)e_{\bf{G}_1}(\bf{k}+\bf{G}_2)e_{\bf{G}_2}(\bf{k}). \nonumber
\end{align}
Since this condition holds for every $\bf{k}$, we may assume, in what concerns the Bloch wave function, $V_{\bf{G}_1+\bf{G}_2}=V_{\bf{G}_1}V_{\bf{G}_2}$, i.e., we have a unitary representation of $\mathbb{Z}^2$. We can then split $V_\mathbf{G}$ into unitary irreducibles, which are parameterized in terms of real unit cell positions $\bf{r}$: $e^{-2\pi i\bf{G}\cdot\bf{r}}$, with $\bf{r}$ and $\bf{r}+\bf{R}$, for $\bf{R}\in\mathbb{Z}^2$, determining the same irreducible representation. This decomposition gives us $u_{\bf{k}}(\bf{r})\in \mathcal{H}_{\bf{r}}$, where $\mathcal{H}_{\bf{r}}$ is a direct summand in the decomposition of the representation into irreducibles and it has the property that for each $\bf{G}\in\mathbb{Z}^2$: $V_{\bf{G}}u_{\bf{k}}(\bf{r})=e^{-2\pi i\bf{G}\cdot \bf{r}}u_{\bf{k}}(\bf{r})$. We note that $u_{\bf{k}}(\bf{r})$ can be a spinor containing multiple components.  If $\mathcal{H}$ is a finite dimensional Hilbert space, $\bf{r}$ takes values in a discrete subset of the real space unit cell, otherwise it may be the whole of the unit cell---the decomposition is then a direct-integral decomposition.  This discussion allows us to derive the relation
\begin{align}
u_{\bf{k}+\bf{G}}(\bf{r})=e_{\bf{G}}(\bf{k})e^{-2\pi i\bf{G}\cdot \bf{r}}u_{\bf{k}}(\bf{r}), 
\label{eq: irrep form of the quasiperiodicity condition}
\end{align}
namely $u_{\bf{k}}(\bf{r})$ behaves as a smooth section of a line bundle $L_{\bf{r}}\to\BZ^2$ whose multipliers are $e_{\bf{G}}(\bf{k})e^{-2\pi i\bf{G}\cdot\bf{r}}$. We point out that $L$ is the ``basic line'' bundle over the Brillouin zone responsible for the topological twist of $\ket{u_{\bf{k}}}$ and the line bundles $L_{\bf{r}}$, all of them topologically isomorphic to $L$, carry information about the real-space unit cell through the variable $\bf{r}$. Unlike earlier works~\cite{wang:cano:millis:liu:yang:21,wang:klevstov:liu:22}, we do not assume spatial periodicity of $u_{\bf{k}}(\bf{r})$, but rather allow a more general quasi-periodicity condition, where $u_{\bf{k}}(\bf{r})$ translated by a lattice vector acquires a phase compatible with a possible net magnetic field present in a unit cell. We see that such a phase factor is necessary to obtain Landau levels, as explicitly derived in Eq.~\eqref{eq: Hilbert bundle multipliers} of Sec.~\ref{sec: technical results}.

Now we fix a translation-invariant complex structure on $\BZ^2$ described by the complex coordinate $z_{\tau}=k_x+\tau k_y$.
The condition for $\ket{u_{\bf{k}}}$ to determine a K\"{a}hler band is~\cite{mera:ozawa:21}
\begin{align}
\frac{\partial}{\partial \bar{z}_{\tau}}\ket{u_{\bf{k}}}-\frac{\bra{u_{\bf{k}}}\frac{\partial}{\partial \bar{z}_{\tau}}\ket{u_{\bf{k}}}}{\bra{u_{\bf{k}}}u_{\bf{k}}\rangle }\ket{u_{\bf{k}}}=0,
\end{align}
which essentially says that $\ket{u_{\bf{k}}}$ is holomorphic up to an overall nonholomorphic nonvanishing multiplicative factor and so, after going to a holomorphic gauge so that the Berry gauge field is represented by a $(1,0)$-form ($\bra{u_{\bf{k}}}\frac{\partial}{\partial \bar{z}_{\tau}}\ket{u_{\bf{k}}}=0$), 
\begin{align}
\frac{\partial}{\partial \bar{z}_{\tau}}\ket{u_{\bf{k}}}=0,
\end{align}
which is simply the requirement of holomorphicity of $\ket{u_{\bf{k}}}$. 

This holomorphicity condition in turn implies, upon appropriate gauge choice, that the multipliers $e_{\bf{G}}(\bf{k})$ are holomorphic in $z_{\tau}$ and that $L\to\BZ^2$ and, as a matter of fact, all of the $L_{\bf{r}}\to\BZ^2$, for $\bf{r}$ running in the real space unit cell, are holomorphic line bundles. Because holomorphic line bundles over complex tori, i.e., $\mathbb{R}^2/\mathbb{Z}^2$ equipped with a translation-invariant complex structure $J$, are determined, up to isomorphism, by the holomorphic line bundles whose spaces of holomorphic sections are described by theta functions~\cite{mumford:07,mumford:74} the form of $\ket{u_{\bf{k}}}$ is already heavily constrained. This is because this condition together with Eq.~\eqref{eq: irrep form of the quasiperiodicity condition} tells us that $u_{\bf{k}}(\bf{r})$ behaves as a holomorphic section of $L_{\bf{r}}$. The space of such sections is a finite dimensional vector space, denoted $H^0(\mathbb{C}/\Lambda_{\tau},L_{\bf{r}})$ and whose dimension equals, by the Riemann-Roch theorem~\cite{miranda:95,huybrechts:05}, $\mathcal{C}=\deg(L_{\bf{r}})$. Moreover, we can describe $H^0(\mathbb{C}/\Lambda_{\tau},L_{\bf{r}})$ quite explicitly in terms of theta functions. Namely, we may assume that~\cite{beauville:13}, after multiplication of $\ket{u_{\bf{k}}}$ by a suitable global nonvanishing holomorphic function $g(\bf{k})$ (in the universal cover $\mathbb{R}^2$)  
\begin{align}
e_{\bf{G}}(\bf{k})= e^{-i\pi \tau \mathcal{C}m_y^2 -2\pi i\mathcal{C} m_y z_{\tau}}, \text{ with } \bf{G}=(m_x,m_y)\in\mathbb{Z}^2,
\end{align}
and then this forces the components of $u_{\bf{k}}(\bf{r})\in\mathcal{H}_{\bf{r}}$, once a basis for $\mathcal{H}_{\bf{r}}$ is chosen, to be linear combinations with possibly $\bf{r}$-dependent coefficients of the theta functions
\begin{align}
\theta_{\bf{r},\alpha}(z_{\tau}):=\thetachar{\frac{\alpha}{\mathcal{C}} -\frac{x}{\mathcal{C}}}{y}(\mathcal{C}z_{\tau},\mathcal{C}\tau),\ \alpha=0,\dots,\mathcal{C}-1,
\end{align}
where $\thetachar{a}{b}(z_{\tau},\tau)=\sum_{n\in\mathbb{Z}} e^{i\pi\tau \left(n+a\right)^2 +2\pi i\left(n+a\right)\left(z_{\tau}+b\right)}$ is known as the theta function with characteristics prescribed by $a,b\in\mathbb{R}$.

{\it Uniqueness of flat K\"{a}hler bands.---} We now require the quantum metric to be flat and derive a much more restrictive condition on $\ket{u_{\bf{k}}}$. Because a K\"{a}hler band is essentially a holomorphic immersion of a complex torus in a projective space, the translation-invariance of the quantum geometry implies that, after fixing one reference quasimomentum, say the zero vector, each translation vector $\bf{k}\in \mathbb{R}^2$ lifts to a quantum symmetry---a symmetry of the target projective space equipped with the Fubini-Study metric---relating $\ket{u_{0}}$ and $\ket{u_{\bf{k}}}$. Once again, due to Wigner's theorem, quantum symmetries are realized by a projective representation of the symmetry group in question, where the transformations are either unitary or anti-unitary.
The key step now is to note that there exist some unitary or anti-unitary operators $U_{\bf{k}}$, $\bf{k}\in\mathbb{R}^2$, such that, in a suitable gauge,
\begin{align}
\ket{u_{\bf{k}}}=U_{\bf{k}}\ket{u_0}, \text{ for } \bf{k}\in\mathbb{R}^2,
\label{eq: coherent state}
\end{align}
which is a consequence of Calabi's rigidity theorem~\cite{calabi:53}; see Sec.~\ref{sec: technical results} for a proof of the above result. We cannot allow $U_\mathbf{k}$ to be anti-unitary, because, if so, the Berry curvature at $\bf{k}$ and $0$ would change sign contradicting the fact that it is assumed to be constant. Hence, we are lead to looking at projective unitary representations of $\mathbb{R}^2$. What this means is that for all $\bf{k}_1,\bf{k}_2\in\mathbb{R}^2$ we have
\begin{align}
U_{\bf{k}_1}U_{\bf{k}_2}=U_{\bf{k}_1+\bf{k}_2}\psi(\bf{k}_1,\bf{k}_2),
\end{align}
for $\psi(\bf{k}_1,\bf{k}_2)$ a $\mathrm{U}(1)$-valued $2$-cocycle, i.e., $\psi(\bf{k}_1,\bf{k}_2)$ are phases satisfying
\begin{align}
\psi(\bf{k}_1,\bf{k}_2+\bf{k}_3)\psi(\bf{k}_2,\bf{k}_3)=\psi(\bf{k}_1,\bf{k}_2)\psi(\bf{k}_1+\bf{k}_2,\bf{k}_3).
\end{align}
We note that $U_\mathbf{G} = V_\mathbf{G}$ for reciprocal lattice vectors $\mathbf{G}$. Projective unitary representations of $\mathbb{R}^2$ are equivalent to certain unitary representations of central extensions $G$ of $\mathbb{R}^2$ by $\mathrm{U}(1)$. Mathematically this means there exists \emph{a short exact sequence of groups}
\begin{align}
1\longrightarrow \mathrm{U}(1) \longrightarrow G \longrightarrow \mathbb{R}^2 \longrightarrow 0,
\end{align}
where each arrow is a group homomorphism and the kernel of each arrow is the image of the previous one. The group $G$ as a set is just the Cartesian product $\mathbb{R}^2\times \mathrm{U}(1)$. As a group, we equip it with the product law
\begin{align}
(\bf{k}_1,\lambda_1)\cdot (\bf{k}_2,\lambda_2)=(\bf{k}_1+\bf{k}_2,\lambda_1\lambda_2 \psi(\bf{k}_1,\bf{k}_2)).
\end{align}
It is then not hard to see that $U(g):=U_{\bf{k}}\lambda$ with $g=(\bf{k},\lambda)\in G$ satisfies $U(g_1)U(g_2)=U(g_1\cdot g_2)$ for all $g_1,g_2\in G$ and hence gives a unitary representation of $G$. It is useful to define the commutator $s(\bf{k}_1,\bf{k}_2)=\psi(\bf{k}_1,\bf{k}_2)/\psi(\bf{k}_2,\bf{k}_1)$ which satisfies, see Ref.~\cite{freed:moore:segal:07:2},
\begin{enumerate}[label={(\roman*)},labelindent=-6pt,leftmargin=!]
    \item Antisymmetry: $s(\bf{k}_1,\bf{k}_2)=s^{-1}(\bf{k}_2,\bf{k}_1)$
    \item Alternating: $s(\bf{k},\bf{k})=1$
    \item Bimultiplicativity: $s(\bf{k}_1\!+\!\bf{k}_2,\bf{k}_3) =s(\bf{k}_1,\bf{k}_3)s(\bf{k}_2,\bf{k}_3)$ \\
    \phantom{Bimultiplicativity}$s(\bf{k}_1,\bf{k}_2\!+\!\bf{k}_3) =s(\bf{k}_1,\bf{k}_2)s(\bf{k}_1,\bf{k}_3)$.
\end{enumerate}
The three properties imply existence of an antisymmetric bilinear form $\omega$ in $\mathbb{R}^2$ with the property $s(\bf{k}_1,\bf{k}_2)=e^{i\omega(\bf{k}_1,\bf{k}_2)}$, where $\omega(\bf{k}_1,\bf{k}_2):=D \bf{k}_1\times \bf{k}_2$, for some $D\in\mathbb{R}$ and $\bf{k}_1\times \bf{k}_2=\bf{k}^t\left(\begin{array}{cc}
   0  & 1 \\
  -1   &  0
\end{array}\right)\bf{k}_2=k_{1,x}k_{2,y}-k_{1,y}k_{2,x}$. 

Theorem~1 of Ref.~\cite{freed:moore:segal:07:2} (see also Refs.~\cite{mumford:07,freed:moore:segal:07:1} for more on Heisenberg groups and central extensions) implies that $G$ is, up to isomorphism, uniquely determined by $s(\bf{k}_1,\bf{k}_2)$. When $\omega$ is non-degenerate, i.e., for $D\neq 0$, then $G$ is referred to as a \emph{Heisenberg group}. A particular realization of $G$ for a given $D$ is determined by setting
\begin{align}
\psi(\bf{k}_1,\bf{k}_2)=e^{iDk_{1,x}k_{2,y}}.
\end{align}
All other realizations are obtained in terms of non-vanishing $\mathrm{U}(1)$-valued functions $g(\bf{k})$ as $\psi'(\bf{k}_1,\bf{k}_2)=\frac{g(\bf{k}_1+\bf{k}_2)}{g(\bf{k}_1)g(\bf{k}_2)}\psi(\bf{k}_1,\bf{k}_2)$ (note that $s(\bf{k}_1,\bf{k}_2)$ is invariant under this change). Observe that with this realization of the central extension $G$, $D=0$ corresponds to a trivial central extension where $G$ is really the direct product of groups $\mathbb{R}^2\times \mathrm{U}(1)$. This last case will not be relevant to us as we shall see below, because $\mathcal{C}\neq 0$.

Now the \emph{Stone-von Neumann theorem}~\cite{mumford:07}, states that the Heisenberg group $G$ has, up to unitary isomorphism, a unique unitary irreducible representation for which $\mathrm{U}(1)$ acts as $(0,\lambda)\cdot \psi=\lambda \cdot \psi$, $(0,\lambda)\in\mathrm{U}(1)\subset G$ and $\psi\in\mathcal{H}$. We will later see that this unitary degree of freedom corresponds to unitary gauge transformations. This irrep is the familiar Hilbert space of square-integrable functions $\mathcal{H}=L^2(\mathbb{R})$ of one variable $q$, where we have the commutation relation $[q,p]=i/D$, with $p=\frac{1}{i D}\frac{\partial}{\partial q}$. At this point, the variable $q$ describes an abstract coordinate, but we derive how one can transform from $q$ to the more physical position coordinate $\mathbf{r}$ in Sec.~\ref{sec: technical results}.
Then,
\begin{align}
U_{\bf{k}}\psi(q) &=e^{i D k_y p}e^{-i D k_x x}\psi(q)\nonumber \\
&=e^{-iD k_x \left(q+k_y\right)}\psi(q+k_y).
\end{align}
Indeed, it is easy to see that $U_{\bf{k}_1}U_{\bf{k}_2}=U_{\bf{k}_1+\bf{k}_2}\psi(\bf{k}_1,\bf{k}_2)$. Note that $D$ takes, in $L^2(\mathbb{R})$, the role of the inverse of Planck's constant.
We note that it is this step which requires that the number of bands, which is the dimension of the Hilbert space $\mathcal{H}$, must be infinite.

We can write, using the Baker-Campbell-Hausdorff formula,
\begin{align}
U_{\bf{k}}= e^{i D\left(k_y p - k_x q\right)}e^{\frac{i}{2} D k_x k_y}.
\end{align}
Using $z_{\tau}=k_x+\tau k_y$, we can then write
\begin{align}
i\left(k_y p - k_x q\right)=\frac{1}{2\tau_2 }\left[z_{\tau} \left(\bar{\tau}q+p\right)-\bar{z}_{\tau}\left(\tau q+p\right)\right]. 
\end{align}
Now observe that, because $q$ and $p$ are self-adjoint, $\bar{\tau}q+p$ is the adjoint to $\tau q+p$ and that 
\begin{align}
[\bar{\tau}q+p,\tau q+p]=i\frac{\bar{\tau}}{D}-i \frac{\tau}{D} =\frac{2\tau_2}{D}.  
\end{align}
It is then convenient to introduce bosonic creation and annihilation operators 
\begin{align}
a_{\tau}:=\sqrt{\frac{D}{2\tau_2}}\left(\tau q+ p\right) \text{ and } a_{\tau}^\dagger:=\sqrt{\frac{D}{2\tau_2}}\left(\bar{\tau} q+ p\right),
\end{align}
which satisfy the canonical commutation relations $[a_{\tau},a_{\tau}^\dagger]=1$ and we may write
\begin{align}
U_{\bf{k}} &=e^{i\frac{D}{2} k_x k_y}e^{\frac{1}{\sqrt{2 D\tau_2}}\left(z_{\tau} a_{\tau}^\dagger -\bar{z}_{\tau} a_{\tau}\right)} \nonumber \\
&=e^{i\frac{D}{2} k_x k_y}e^{-\frac{1}{4\tau_2 D}|z_{\tau}|^2}e^{\frac{1}{\sqrt{2 D\tau_2}} z_{\tau} a_{\tau}^\dagger}e^{-\frac{1}{\sqrt{2 D\tau_2}}\bar{z}_{\tau} a_{\tau}}.
\end{align}
It is then clear that, after changing gauge, canceling all nonholomorphic nonvanishing multiplicative factors, we see that to have a holomorphic $\ket{u_{\bf{k}}}$ we must have $\frac{\partial}{\partial \bar{z}_{\tau}}\left(e^{-\frac{1}{\sqrt{2 D\tau_2}}\bar{z}_{\tau} a_{\tau}}\ket{u_0}\right)=0\iff a_{\tau}\ket{u_0}=0$. Now the condition $a_{\tau}\ket{u_0}=0$ is just the statement that $\ket{u_0}$ is the groundstate of the bosonic mode $a_{\tau}$. The only solution, up to an overall constant, is
\begin{align}
u_0(q)= e^{-\frac{i D \tau}{2} q^2},  
\end{align}
which is in $L^2(\mathbb{R})$ iff $D<0$. So a holomorphic state exists iff $D<0$. This does not preclude us from considering other representations of $G$ that are not irreducible. However, due to the above finding that $\ket{u_0}\in \mathcal{H}$ is unique (up to rescale) this actually implies that the resulting K\"{a}hler band is unique in a precise sense. The reason is as follows. Suppose $U_{\bf{k}}$ was reducible. Then the Hilbert space assumes the form $\mathcal{H}\oplus \dots \oplus \mathcal{H}\cong \mathcal{H}\otimes \mathbb{C}^{l}$, where $\mathcal{H}$ is the unique irrep of $G$ and $l$ is the number of times it appears ($l$ may not be finite, but this does not change the argument). The group $G$ then acts only on the left factor in $\mathcal{H}\otimes \mathbb{C}^l$. Due to there existing only one possible choice (up to rescale) for $\ket{u_0}$ in $\mathcal{H}$ that we can take in Eq.~\eqref{eq: coherent state}, this then implies that the augmented state in $\mathcal{H}\otimes \mathbb{C}^l$ will necessarily be of the form $\ket{u_{\bf{k}}}\otimes \ket{c}$ for some constant nonzero vector $\ket{c}\in\mathbb{C}^l$ and $\ket{u_{\bf{k}}}\in\mathcal{H}$ as in Eq.~\eqref{eq: coherent state}. This just corresponds to taking the same Bloch band and introducing a new degree of freedom, such as spin, which decouples and hence the resulting Bloch band has a definite ``spin polarization.''

Finally, we use the condition Eq.~(\ref{eq:quasiperiodic}) to find the value of $D$.
Note that
\begin{align}
\ket{u_{\bf{k}+\bf{G}}}=\psi^{-1}(\bf{G},\bf{k}) U_{\bf{G}}U_{\bf{k}}\ket{u_{0}}=\psi^{-1}(\bf{G},\bf{k}) U_{\bf{G}}\ket{u_{\bf{k}}}.
\end{align}
Then, $e_{\bf{G}}(\bf{k}):=\psi^{-1}(\bf{G},\bf{k})$ should be unitary (since $U_{\bf{k}}$ is unitary) multipliers for a line bundle over $\BZ^2$, and thus
\begin{align}
e_{\bf{G}}(\bf{k})=e^{-i D m_x k_y} \text{ with } \bf{G}=(m_x,m_y)\in\mathbb{Z}^2.
\end{align}
A connection on the line bundle $L$ is determined by a one-form that is compatible with the system of multipliers:
\begin{align}
A(\bf{k}+\bf{G})=A(\bf{k})+ e_{\bf{G}}(\bf{k})de^{-1}_{\bf{G}}(\bf{k})= A(\bf{k}) +i D m_x dk_y.
\end{align}
One possible choice is $A=i D k_x dk_y$. We then see that the Chern number of $L$ being equal to $\mathcal{C}$ forces $D=-2\pi \mathcal{C}$. This concludes the proof of the Theorem, as the K\"{a}hler band we are looking for is precisely $\ket{u_{\bf{k}}}=U_{\bf{k}}\ket{u_0}$ for the Heisenberg group determined by the Chern number $\mathcal{C}$ and $\ket{u_0}$ such that $a_{\tau}\ket{u_0}=0$. Please note how $\mathcal{C}$ enters in $G$ and fixes the $U_{\bf{k}}$'s and $\mathcal{H}$, $\tau$ fixes what $\ket{u_0}$ must be.

We now proceed to give an explicit expression of $u_{\bf{k}}(\bf{r})$ in terms of theta functions---hence using the particular realization of $G$. We refer the reader to Sec.~\ref{sec: technical results}, where we show that $\mathcal{H}_{\bf{r}}\cong \mathbb{C}^{\mathcal{C}}$, corresponding to the internal ``color'' degree of freedom, and
\begin{align}
u_{\bf{k}}(\bf{r})=e^{2\pi i \mathcal{C} k_x k_y} e^{i\pi\mathcal{C} \tau k_y^2}\left(\theta_{\bf{r},0}(z_{\tau}),\dots,\theta_{\bf{r},\mathcal{C}-1}(z_{\tau}) \right).
\label{eq: LLL color-entangled wave function}
\end{align}
For the particular case $\mathcal{C}=1$ and $\tau = i$, this is exactly the lowest Landau level Bloch wave function in the Landau gauge---see Sec.~\ref{sec: technical results} for a detailed derivation of this fact, while for higher $\mathcal{C}$ is the color-entangled lowest Landau level Bloch wave function previously considered in the literature~\cite{wu:regnault:bernevig:13,wang:klevstov:liu:22}.

We remark that the fact that the unitary irrep of $G$ is unique up to unitary isomorphism corresponds to the freedom of gauge choice. Indeed, since the isomorphism intertwines the action of $G$, it must, in particular, preserve the quantum number associated to reciprocal lattice translations $\bf{r}$. It implies that we are allowed to perform $\mathrm{U}(\mathcal{C})$-gauge transformations to $u_\mathbf{k}(\mathbf{r})$. We note that this is also equivalent to choosing a different orthogonal basis of theta functions consistent with the Chern number $\mathcal{C}$ and the irrep of the reciprocal lattice associated with $\bf{r}$. In summary, in Eq.\eqref{eq: LLL color-entangled wave function}, we have two kinds of gauge degrees of freedom:
\begin{itemize}
    \item [(i)] $\mathrm{U}(\mathcal{C})$ real space gauge transformations $u_{\bf{k}}(\bf{r})\mapsto S(\bf{r})u_{\bf{k}}(\bf{r})$, for $S(\bf{r})\in\mathrm{U}(\mathcal{C})$;
    \item [(ii)] $\mathbb{C}^*$ momentum space gauge transformations $u_{\bf{k}}(\bf{r})\mapsto g(\bf{k})u_{\bf{k}}(\bf{r})$, for $g(\bf{k})\in\mathbb{C}^*$.
\end{itemize}
In more physical terms, the transformation (i) corresponds, for example in the case of lowest Landau levels, to go from the Landau gauge to the symmetric gauge, whereas the transformation (ii) corresponds to choosing the normalization and phase of the Bloch wave function at each momentum.
\section{Discussion}
\label{sec: discussion}
We have shown the uniqueness of the geometrically flat K{\"a}hler bands for given Chern number and modular parameter. In previous works, the effect of the modular parameter $\tau$ is little explored; the situation $\tau \neq i$ can arise when the effective mass of a particle is anisotropic, \emph{cf.} Ref.~\cite{haldane:2011} where the Galilean metric includes the information of $\tau$. Our result shows that there is a continuous family of geometrically flat K{\"a}hler bands parameterized by $\tau$, which provides novel degrees of freedom to explore fractional topological physics with flat K{\"a}hler bands. 

It is known that any holomorphic wave function with a constant $J$, known also as ideal Chern bands or ideal K\"ahler bands, can be obtained by modulating geometrically flat Kähler bands~\cite{wang:klevstov:liu:22}. The GMP algebra of density operators crucial in obtaining stable Abelian fractional quantum Hall phases has been shown to be recovered for such ideal K\"ahler bands. A tower of higher Landau level analogs, which can support non-Abelian fractionalized states, has also be constructed starting from ideal K\"ahler bands~\cite{liu:mera:fujimoto:ozawa:wang:2024}. Thus, the uniqueness of the geometrically flat K\"ahler bands proved in this work allows one to write down all possible wave functions which serve as building blocks for these fractional quantum Hall phases.
\section{Technical Results}
\label{sec: technical results}
\subsection{Proof of existence of the projective unitary representation of the translation group under the flatness assumption}
\label{subsec: Proof of existence of the projective unitary representation of the translation group under the flatness assumption}
We wish to proof existence of $U_{\bf{q}}$ as in Eq.~(12) in the main text. The proof is based on Calabi's rigidity theorem. As we will see, we can only determine the action of $U_{\bf{q}}$ on the minimal linear subspace (subspace obtained by projectivization of a vector subspace of $\mathbb{C}^{n+1}$) of $\mathbb{C}P^n$ which contains the immersed submanifold determined by $\ket{u_{\bf{k}}}$ (concretely, $\mathrm{span}\{\ket{u_{\bf{k}}}:\bf{k}\in\mathbb{R}^2\}\subset \mathbb{C}^{n+1}$). But this linear subspace of $\mathbb{C}P^n$ is generally much larger than just the individual one-dimensional subspaces associated to the occupied bands. (For example, for a Chern insulator, even if we consider one band with nonzero Chern number to be occupied, we should take multiple bands whose sum of the Chern numbers is zero.) Thus, $U_{\bf{q}}$ determines a projective representation once we restrict our Bloch wave function so that $\mathbb{C}P^n$ in the target coincides with the minimal linear subspace we need to consider. In physical terms, if we add ``trivial unoccupied bands,'' we cannot determine how $U_{\bf{q}}$ acts on these bands, which is very reasonable, and we restrict ourselves to the situation where such trivial bands are excluded.
\setcounter{theorem}{1}
\begin{theorem}[Rigidity theorem (Calabi)~\cite{calabi:53,takeuchi:78}] Let $M$ be a connected K\"{a}hler manifold and let $f:M\to \mathbb{C}P^n$ and $f':M\to \mathbb{C}P^{n'}$ be K\"{a}hler immersions (i.e., holomorphic immersions of $M$ whose K\"{a}hler structure coincides with the induced K\"{a}hler structure) so that their images do not lie in any proper linear subspace of the projective space (i.e., a subspace of the projective space obtained by projectivization of a vector subspace). Then $n=n'$ and there exists a unitary operator $U\in\mathrm{U}(n+1)$ such that $U\circ f =f'$, where, by abuse of notation, we have also denoted by $U$ the induced diffeomorphism in $\mathbb{C}P^{n}$.
\label{th: rigidity}
\end{theorem}
\noindent We remark that the dimension $n$ in the above Theorem~\ref{th: rigidity} and also below can be infinity. The assumption that the images do not lie in any proper linear subspace of the projective space implies that we have excluded benign trivial bands from consideration as we commented above.
\begin{proposition}Suppose we have a a K\"{a}hler immersion $f:\mathbb{R}^2\to \mathbb{C}P^{n}$ with respect to some complex structure $j$ in the plane. Suppose also that the image $f(\mathbb{R}^2)$ does not lie in any proper linear subspace of $\mathbb{C}P^{n}$. Suppose $\phi:\mathbb{R}^2\to\mathbb{R}^2$ is a biholomorphism, with respect to $j$, preserving the pullback under $f$ of the Fubini-Study metric $f^*g_{FS}$, i.e. a holomorphic isometry of $f^*g_{FS}$. Then there exists a unitary operator $U_{\phi}\in\mathrm{U}(n+1)$, unique up to a phase, such that $f\circ \phi= U_{\phi}\circ f$, where we have also denoted by $U_{\phi}$ the induced diffeomorphism in $\mathbb{C}P^n$. 
\label{prop: lifting of phi}
\end{proposition}
\begin{proof} By assumption, the maps $f\circ \phi: \mathbb{R}^2\to\mathbb{C}P^n$ and $f:\mathbb{R}^2\to\mathbb{C}P^n$ are both K\"{a}hler immersions. It follows from Theorem~\ref{th: rigidity} that there exists $U_{\phi}\in\mathrm{U}(n+1)$ such that $f\circ \phi =U_{\phi}\circ f$. We now prove the uniqueness of $U_{\phi}$ up to a phase, following the arguments given in the proof of Theorem~4.3. in Ref.~\cite{nakagawa:76}, which we briefly reproduce here. Let us assume that two unitary matrices $U_{\phi},U'_{\phi}\in\mathrm{U}(n+1)$ satisfy $U_{\phi}\circ f=f\circ \phi$ and $U'_{\phi}\circ f=f\circ \phi$. Then it follows that $\left(\left(U'_{\phi}\right)^{-1}U_{\phi}\right)\circ f= f$. Let us define $U:=\left(U'_{\phi}\right)^{-1}U_{\phi}\in\mathrm{U}(n+1)$ and show that $U$ is just a phase, which implies that $U_{\phi}$ and $U'_{\phi}$ differ only by a phase.

We choose choose an orthogonal basis of $\mathbb{C}^{n+1}$ where $U$ is diagonalized, for which $U$ is given by the diagonal matrix $\mathrm{diag}(\alpha_1,\dots, \alpha_{1}, \dots, \alpha_{r},\dots, \alpha_{r})$ with each eigenvalue $\alpha_{i}$, having, possibly, some multiplicity. Now observe that if $v\in\mathbb{C}^{n+1}$ is an eigenvector of $U$, it follows that the one-dimensional subspace of $\mathbb{C}P^n$ generated by $v$, denoted by $\ell_{v}\in\mathbb{C}P^{n}$, is a fixed point of the induced map $U:\mathbb{C}P^{n}\to\mathbb{C}P^n$, i.e., $U(\ell_v)=\ell_v$.  It follows that the set of fixed points of the isometry $U: \mathbb{C}P^{n}\to\mathbb{C}P^{n}$ is given by the disjoint union $\coprod_{i=1}^r S(\alpha_i)$ with $S(\alpha_i)$ being the image of the eigenspace associated with the eigenvalue $\alpha_i$ under the quotient map $\pi: \mathbb{C}^{n+1}-\{0\}\to \mathbb{C}P^n$, $i=1,\dots, r$. Since in $f(\mathbb{R}^2)$ the map induced by the unitary $U$ is the identity and because $f(\mathbb{R}^2)$ is connected ($\mathbb{R}^2$ is connected and $f$ is continuous), it follows that $f(\mathbb{R}^2)\subset S(\alpha_i)$, for some $i$. However, each $S(\alpha_i)$ is a linear subspace of $\mathbb{C}P^n$ By the assumption that $f(\mathbb{R}^2)$ does not lie in any proper linear subspace of $\mathbb{C}P^{n}$ it follows that $r = 1$ and $S(\alpha_1) = \mathbb{C}P^n$. Therefore, $U$ is just a multiple of the identity by a phase factor, as we wanted to show.
\end{proof}
\begin{theorem} Suppose we have a K\"{a}hler immersion $f:\mathbb{R}^2\to\mathbb{C}P^{n}$ with respect to some complex structure $j$ in the plane. Suppose also that the image $f(\mathbb{R}^2)$ does not lie in any proper linear subspace of $\mathbb{C}P^{n}$. Suppose a group $G$ acts in $\mathbb{R}^2$ by holomorphic isometries, $\phi_{g}:\mathbb{R}^2\to\mathbb{R}^2$ for $g\in G$, of $f^*g_{FS}$. Then there exists a projective unitary representation $U:G\to\mathrm{PU}(n+1)$ with the property $f\circ \phi_g= U(g)\circ f$.
\label{th: projective rep}
\end{theorem}
\begin{proof} For each $g\in G$, $\phi_g$ is a holomorphic isometry of $f^*g_{FS}$. Therefore, by Proposition~\ref{prop: lifting of phi} there exists a $U_g:=U_{\phi_g}\in \mathrm{U}(n+1)$ uniquely defined up to a phase such that $f\circ \phi_g= U_g\circ f$. Furthermore, by assumption, we have $\phi_{g_1g_2}=\phi_{g_1}\circ \phi_{g_2}$ and this implies
\begin{align}
\left(U_{g_1}\circ U_{g_2}\right)\circ f = U_{g_1g_2}\circ f.
\end{align}
The previous equation implies, using the same arguments as in the proof of Proposition~\ref{prop: lifting of phi} to show uniqueness of $U_{\phi}$ up to phase, that $U_{g_1}U_{g_2}=U_{g_1g_2}$ holds projectively, implying that the assignment $g\mapsto U_g$ determines a projective unitary representation of $G$.
\end{proof}
Letting $f:\mathbb{R}^2\to\mathbb{C}P^n$ be the map induced by the Bloch wave function $\ket{u_{\bf{k}}}$, letting $G$ be the translation group $\mathbb{R}^2$ which acts by holomorphic isometries in $\mathbb{R}^2$ equipped with a translation-invariant metric and complex structure---translation-invariant quantum geometry---, the existence of the projective representation $U_\mathbf{q}$ follows from Theorem~\ref{th: projective rep}.

The discussion above does not give a concrete form of the $U_{\bf{q}}$'s, rather it just guarantees their existence. The projective representation is determined up to isomorphism of projective representations. Essentially this means that one can take $U'_{\bf{q}}=U_{\bf{q}} g^{-1}(\bf{q})$ , with $g(\bf{q})$ a phase factor---because this does not change $\langle u_{\bf{k}+\bf{q}}|u_{\bf{k}+\bf{q}}\rangle$. This means that if
\begin{align}
U_{\bf{q}_1}U_{\bf{q}_2}=U_{\bf{q}_1+\bf{q}_2} \psi(\bf{q}_1,\bf{q}_2), 
\end{align}
then
\begin{align}
U'_{\bf{q}_1}U'_{\bf{q}_2}=U'_{\bf{q}_1+\bf{q}_2} \psi(\bf{q}_1,\bf{q}_2)\frac{g(\bf{q}_1+\bf{q}_2)}{g(\bf{q}_1)g(\bf{q}_2)},
\end{align}
and we see that $\psi(\bf{q}_1,\bf{q}_2)$ and $\psi(\bf{q}_1,\bf{q}_2)\frac{g(\bf{q}_1+\bf{q}_2)}{g(\bf{q}_1)g(\bf{q}_2)}$ differ by an exact cocycle. Later in the maintext, it is found that this cocycle is canonically specified by the Chern number $\mathcal{C}$.

The explicit construction of $U_{\bf{q}}$ then comes from the argument of uniqueness of the representation (where the phase factors act in the standard way) of the central extension of the translation group by $\mathrm{U}(1)$ which is the content of the Stone-von Neumann theorem. 

\subsection{Obtaining the lowest Landau level type wave functions from an explicit realization of  the Heisenberg group $G$}
\label{sec: Obtaining the lowest Landau level type wave functions from an explicit realization of  the Heisenberg group G}
We derive an explicit expression of $u_{\bf{k}}(\bf{r})$ in terms of theta functions---this will allow us to recover the lowest Landau level wave function and the color-entangled wave functions. To do this, we need to extract the component $u_{\bf{k}}(\bf{r})$ corresponding to the irreducible representation of $\mathbb{Z}^2$ labeled by $\bf{r}$. Hence, the defining property of $u_{\bf{k}}(\bf{r})$ is $U_{\bf{G}}u_{\bf{k}}(\bf{r})=e^{-2\pi i\bf{G}\cdot\bf{r}}u_{\bf{k}}(\bf{r})$ for all $\bf{G}\in\mathbb{Z}^2$. This can be done through the \emph{Bloch-Zak transform} which takes $\ket{u_{\bf{k}}}\in \mathcal{H}=L^2(\mathbb{R})$ to
\begin{align}
u_{\bf{k}}(\bf{r})=\sum_{\bf{G}\in\mathbb{Z}^2}e^{2\pi i \bf{r}\cdot \bf{G}} U_{\bf{G}} \ket{u_{\bf{k}}},
\end{align}
and provides a Hilbert space isomorphism $L^2(\mathbb{R})\cong \directint_{\text{u.c.}}d^2\bf{r}\; \mathcal{H}_{\bf{r}}$ where $\text{u.c.}$ stands for the real space unit cell labeling irreps of the reciprocal lattice $\mathbb{Z}^2$. For the transform to be an isomorphism we need to define the inner product in $\mathcal{H}_{\bf{r}}$ appropriately. For a general element $\ket{f}\in L^2(\mathbb{R})$, let us denote its value at a point $q \in \mathbb{R}$ by $f(q)$. The Bloch-Zak transform determines an element $f(\bf{r})\in\mathcal{H}_{\bf{r}}$ from $\ket{f}\in L^2(\mathbb{R})$, which evaluated at a point $q \in \mathbb{R}$ reads 
\begin{align}
&\left(f(\bf{r})\right)(q) =\sum_{\bf{G}\in\mathbb{Z}^2} e^{2\pi i\bf{r}\cdot\bf{G}} U_{\bf{G}} f(q) \notag \\
&=\sum_{m_x,m_y\in\mathbb{Z}^2} e^{2\pi i\mathcal{C}m_x\left( q+m_y\right) +2\pi i xm_x +2\pi i ym_y}f(q+m_y) \notag \\
&=
\sum_{m_x,m_y\in\mathbb{Z}^2} e^{2\pi i(\mathcal{C}q + x)m_x + 2\pi i ym_y}f(q+m_y),
\end{align}
where we wrote $\mathbf{G} = (m_x, m_y)$ and $\mathbf{r} = (x,y)$ and used Eq.(18) of the main text. One can easily check that the inverse transformation is $\ket{f}=\int_{\text{u.c.}}d^2\bf{r}\; f(\bf{r})$.
Using an expression of the series expansion of the Dirac comb, $\sum_{p \in \mathbb{Z}}\delta (t -p) = \sum_{m\in \mathbb{Z}}e^{2\pi i t m}$, we obtain
\begin{align}
&\left(f(\bf{r})\right)(q) 
    = \sum_{m_y\in\mathbb{Z}}\sum_{p\in \mathbb{Z}} \delta (\mathcal{C}q + x - p)e^{2\pi i ym_y}f(q+m_y)
    \notag \\
    &=
    \sum_{m_y\in\mathbb{Z}}\sum_{p\in \mathbb{Z}} \delta (\mathcal{C}q + x - p)e^{2\pi i ym_y}f\left( -\frac{x}{\mathcal{C}} + \frac{p}{\mathcal{C}}+m_y \right).
\end{align}
We now write $p = \alpha + \tilde{p} \mathcal{C}$ where $\alpha\in \{0,\dots ,\mathcal{C}-1\}$ and $\tilde{p}\in\mathbb{Z}$ so that $\sum_{ p \in \mathbb{Z}} = \sum_{\alpha = 0}^{\mathcal{C}-1}\sum_{\tilde{p}\in \mathbb{Z}}$. Then,
\begin{widetext}
\begin{align}
\left(f(\bf{r})\right)(q) 
    &=
    \sum_{m_y \in \mathbb{Z}}\sum_{\alpha = 0}^{\mathcal{C}-1}\sum_{\tilde{p}\in \mathbb{Z}} \delta (\mathcal{C}q + x - \alpha - \mathcal{C}\tilde{p})e^{2\pi i ym_y}f\left( -\frac{x}{\mathcal{C}} + \frac{\alpha}{\mathcal{C}}+m_y + \tilde{p} \right)
    \notag \\
    &=
    \sum_{m_y \in \mathbb{Z}}\sum_{\alpha = 0}^{\mathcal{C}-1}\sum_{\tilde{p}\in \mathbb{Z}} \delta (\mathcal{C}q + x - \alpha - \mathcal{C}\tilde{p})e^{2\pi i y(m_y-\tilde{p})}f\left( -\frac{x}{\mathcal{C}} + \frac{\alpha}{\mathcal{C}}+m_y \right)
    \notag \\
    &=\sum_{\alpha=0}^{\mathcal{C}-1}\left(\sum_{m_y\in\mathbb{Z}} f\left(-\frac{x}{\mathcal{C}}+\frac{\alpha}{\mathcal{C}}+m_y\right) e^{2\pi i\frac{1}{\mathcal{C}} y \left(-x+\alpha + \mathcal{C} m_y\right)}\right)\left(\sum_{\tilde{p}\in\mathbb{Z}} \delta(\mathcal{C}q +x-\alpha -\mathcal{C}\tilde{p}) e^{-2\pi i \frac{1}{\mathcal{C}}y \left(-x+\alpha + \mathcal{C}\tilde{p}\right)}\right)
    \notag \\
    &\equiv
    \sum_{\alpha=0}^{\mathcal{C}-1} f_\alpha (\mathbf{r}) \delta_\alpha^\mathbf{r}(q),
\end{align}
\end{widetext}
where we defined
\begin{align}
\delta_{\alpha}^{\bf{r}}(q):&=\sum_{p\in\mathbb{Z}} \delta(\mathcal{C}q +x-\alpha -\mathcal{C}p) e^{-2\pi i \frac{1}{\mathcal{C}}y \left(-x+\alpha + \mathcal{C}p\right)},
\\
f_{\alpha}(\bf{r}):&=\sum_{m_y\in\mathbb{Z}} f\left(-\frac{x}{\mathcal{C}}+\frac{\alpha}{\mathcal{C}}+m_y\right) e^{2\pi i\frac{1}{\mathcal{C}} y \left(-x+\alpha + \mathcal{C} m_y\right)}.
\end{align}
We can regard the vector space $\mathcal{H}_{\bf{r}}$ to be the span of the distributions $\delta_{\alpha}^{\bf{r}}(q)$.
We see that periodicity $f(\mathbf{r}+\mathbf{R}) = f(\mathbf{r})$ holds because $\delta_{\alpha}^{\bf{r}}(q)$ and $f_{\alpha}(\bf{r})$ transform in the opposite ways; for $\bf{R}=(a_x,a_y)\in\mathbb{Z}^2$,
\begin{widetext}
\begin{align}
\delta_{\alpha}^{\bf{r}+\bf{R}}(q)=e^{-2\pi i \frac{1}{\mathcal{C}}a_y \left(-x+a_x+\alpha\right)}\delta^{\bf{r}}_{\alpha - a_x}(q) \text{ and } \delta_{\alpha+\mathcal{C}}^{\bf{r}}=\delta_{\alpha}^{\bf{r}},\ \alpha=0,\dots,\mathcal{C}-1, \\
    f_{\alpha}(\bf{r}+\bf{R})=e^{2\pi i \frac{1}{\mathcal{C}}a_y \left(-x+a_x+\alpha\right)}f_{\alpha - a_x}(\bf{r}) \text{ and } f_{\alpha+\mathcal{C}}(\bf{r})=f_{\alpha}(\bf{r}),\ \alpha=0,\dots,\mathcal{C}-1.
\label{eq: Hilbert bundle multipliers}
\end{align}
\end{widetext}
This periodicity is consistent with the requirement of the periodicity of the Hilbert space, $\mathcal{H}_{\bf{r}+\bf{R}}=\mathcal{H}_{\bf{r}}$ coming from $e^{-2\pi i\bf{G}\cdot\left(\bf{r}+\bf{R}\right)}=e^{-2\pi i\bf{G}\cdot\bf{r}}$, which holds for any $\bf{G}$ in the reciprocal lattice.

If we define a Hilbert space structure by declaring
\begin{align}
\langle \delta_{\alpha}^{\bf{r}},\delta_{\beta}^{\bf{r}}\rangle =\frac{1}{\mathcal{C}} \delta_{\alpha\beta},
\end{align}
then we get the desired Hilbert space isomorphism  $L^2(\mathbb{R})\cong \directint_{\text{u.c.}}d^2\bf{r}\; \mathcal{H}_{\bf{r}}$, where the right-hand side can be interpreted as the space of square-integrable sections of a Hilbert bundle over the real space unit cell torus, obtained by identifying $\bf{r}\sim \bf{r}+\bf{R}$ with $\bf{R}\in\mathbb{Z}^2$, and whose fiber is $\mathcal{H}_{\bf{r}}$. A section of this Hilbert bundle is identified with a collection of smooth functions $f_0(\bf{r}),\dots, f_{\mathcal{C}-1}(\bf{r})$ satisfying the conditions in Eq.~\eqref{eq: Hilbert bundle multipliers}. The collection $\{\delta_{\alpha}^{\bf{r}}\}_{\alpha=0}^{\mathcal{C}-1}$ can then be interpreted as a global multivalued orthogonal frame field for this Hilbert bundle. To see that it is indeed a Hilbert space isomorphism, one shows that for $f,g\in L^2(\mathbb{R})$, we have
\begin{widetext}
\begin{align}
&\frac{1}{\mathcal{C}}\sum_{\alpha=0}^{\mathcal{C}-1}\int_{\text{u.c.}}d^2\bf{r}\; \overline{f_{\alpha}(\bf{r})}g_{\alpha}(\bf{r}) \nonumber \\
&=\frac{1}{\mathcal{C}}\sum_{\alpha=0}^{\mathcal{C}-1}\int_{0}^{1}dx\; \sum_{m_y\in\mathbb{Z}}\overline{f\left(-\frac{x}{\mathcal{C}}+\frac{\alpha}{\mathcal{C}}+m_y\right)}g\left(-\frac{x}{\mathcal{C}}+\frac{\alpha}{\mathcal{C}}+m_y\right)\nonumber \\
&=\int_{\mathbb{R}}dq\; \overline{f(q)}g(q)=\langle f|g\rangle,
\end{align}
\end{widetext}
where we have used $\int_{0}^1dy\; e^{2\pi i y(m-n)}=\delta_{m,n}$ for $m,n\in\mathbb{Z}$. We also refer to $(f_0(\bf{r}),\dots, f_{\mathcal{C}-1}(\bf{r}))$ as the Bloch-Zak transform of $\ket{f}$---we are explicitly using the isomorphism $\mathcal{H}_{\bf{r}}\cong \mathbb{C}^{\mathcal{C}}$ provided by the orthogonal basis determined by the $\delta_{\alpha}^{\bf{r}}$'s. Physically, these $\mathcal{C}$ degrees of freedom can be spins, orbitals, or other internal degrees of freedom.

Now we explicitly evaluate the Bloch-Zak transform for $\ket{f}=\ket{u_{\bf{k}}}$. First note that, From Eqs.(18) and (24) of the main text,
\begin{align}
u_{\bf{k}}(q)=U_{\bf{k}}u_0(q)=e^{i\pi\mathcal{C}\tau\left(q+k_y\right)^2 +2\pi i\mathcal{C}k_x\left(q+k_y\right)}. \label{eq:uk}
\end{align}
Then we find,
\begin{widetext}
\begin{align}
f_{\alpha}(\bf{r}) &= \sum_{m_y\in\mathbb{Z}} u_\mathbf{k}\left(-\frac{x}{\mathcal{C}}+\frac{\alpha}{\mathcal{C}}+m_y\right) e^{2\pi i\frac{1}{\mathcal{C}} y \left(-x+\alpha + \mathcal{C} m_y\right)}
 \\
&=\sum_{m_y\in\mathbb{Z}} e^{i\pi\mathcal{C}\tau\left(-\frac{x}{\mathcal{C}}+\frac{\alpha}{\mathcal{C}}+k_y+m_y\right)^2 +2\pi i \mathcal{C} k_x\left(-\frac{x}{\mathcal{C}}+\frac{\alpha}{\mathcal{C}}+k_y+m_y\right)+2\pi i\frac{1}{\mathcal{C}}y\left(-x+\alpha+\mathcal{C}m_y\right)} \nonumber \\
&= e^{2\pi i \mathcal{C} k_x k_y} e^{i\pi\mathcal{C} \tau k_y^2}\thetachar{\frac{\alpha}{\mathcal{C}} -\frac{x}{\mathcal{C}}}{y}(\mathcal{C}z_{\tau},\mathcal{C}\tau),
\end{align}
\end{widetext}
where $z_\tau = k_x + \tau k_y$, and we have introduced the theta functions with characteristics $a,b\in\mathbb{R}$ as 
\begin{align}
\thetachar{a}{b}(z,\tau):=\sum_{n\in\mathbb{Z}}e^{i\pi\tau(n+a)^2 +2\pi i(n+a)(z+b)},
\end{align}
for $z \in \mathbb{C}$.
Therefore, the real-space representation of the Bloch state $\ket{u_\mathbf{k}}$ is thus given by a $\mathcal{C}$-component wave function of the following form:
\begin{widetext}
\begin{align}
u_{\bf{k}}(\bf{r})=e^{2\pi i \mathcal{C} k_x k_y} e^{i\pi\mathcal{C} \tau k_y^2}\left(
\thetachar{-\frac{x}{\mathcal{C}}}{y}(\mathcal{C}z_{\tau},\mathcal{C}\tau),
\dots,
\thetachar{\frac{\mathcal{C} - 1}{\mathcal{C}} -\frac{x}{\mathcal{C}}}{y}(\mathcal{C}z_{\tau},\mathcal{C}\tau)
\right).
\label{eq: LLL Bloch wave functions}
\end{align}    
\end{widetext}
Since the overall normalization of the wave function can be chosen arbitrarily, the $\mathbf{k}$-dependent exponential factors in front of the theta functions can be removed if one wishes. This degree of freedom corresponds to choosing $U_\mathbf{k}$ and the normalization of $u_0 (q)$ different from ones we used in Eq.(\ref{eq:uk}). Furthermore, as noted in the main text, there is a $U(\mathcal{C})$ real-space gauge degree freedom; this is to multiply $u_\mathbf{k}(\mathbf{r})$ by an element of $U(\mathcal{C})$, which can depend on $\mathbf{r}$.

For the particular case $\mathcal{C}=1$ and $\tau = i$, the wave function $u_\mathbf{k}(\mathbf{r})$ is just the lowest Landau level Bloch wave function (in the Landau gauge), while for higher $\mathcal{C}$ is the so-called color-entangled lowest Landau level Bloch wave function. Since the Landau level wave function expressed in terms of theta functions may not be familiar to some readers, we include, for completeness, its derivation in the Appendix. 

\section*{Acknowledgments}
B.~M. is very grateful to J. P. Nunes, J. M. Mour\~{a}o, and T. Baier for very fruitful discussions. This work is supported by JSPS KAKENHI Grant No. JP20H01845, JST PRESTO Grant No. JPMJPR19L2, and JST CREST Grant No. JPMJCR19T1. B.~M. acknowledges support from the Security and Quantum Information Group (SQIG) in Instituto de Telecomunica\c{c}\~{o}es, Lisbon. This work is funded by FCT (Funda\c{c}\~{a}o para a Ci\^{e}ncia e a Tecnologia) through national funds FCT I.P. and, when eligible, by COMPETE 2020 FEDER funds, under Award UIDB/50008/2020 and the Scientific Employment Stimulus---Individual Call (CEEC Individual)---2022.05522.CEECIND/CP1716/CT0001, with DOI 10.54499/2022.05522.CEECIND/CP1716/CT0001.
\bibliography{bib.bib}
\begin{widetext}
\appendix
\section{Lowest Landau level Bloch wave function}
\label{sec: lowest Landau level Bloch wave function}
Here, for completeness, we present a derivation of the lowest Landau level Bloch wave function. The quantity $\bf{r}=(x,y)$ will denote coordinates in the plane. The Hamiltonian of an electron in a uniform magnetic field is given by
\begin{align}
H=-\frac{1}{2}\left[(\partial_x +2\pi i y)^2 + \partial^2_y \right],
\label{eq: Hamiltonian}
\end{align}
where we assume the standard Euclidean metric in the plane and we use (rescaled) coordinates and a (unitary) Landau gauge such that $A=2\pi i y dx$. Note that we use a convention for which the gauge field is purely imaginary (instead of purely real).

The Hamiltonian $H$ is not invariant under the standard action of the translation group of the plane. However, we may introduce a projective action of the translation group such that $H$ is invariant under it. Taking an element $\bf{t}=(t_x,t_y)\in\mathbb{R}^2$, its action on a wave function is given by a unitary operator $U(\bf{t})$, which we refer to as a \emph{magnetic translation}, defined by
\begin{align}
U(\bf{t})\psi(\bf{r})=e^{2\pi i t_y x} \psi(\bf{r}+\bf{t}).    
\end{align}
The action is projective because
\begin{align}
U(\bf{t}_1)U(\bf{t}_2) = U(\bf{t}_1+\bf{t}_2)e^{2\pi i t_{1,x}t_{2,y}}
\label{eq: Heisenberg group}
\end{align}
The set of operators $\{U(\bf{t})\lambda: \bf{t}\in\mathbb{R}^2,\ \lambda \in\text{U}(1)\}$ forms a Heisenberg group, centrally extending the translation group $\mathbb{R}^2$ of the plane by $\mathrm{U}(1)$.

The covariant derivative $\nabla=d+A$, where $d=dx\partial_x +dy\partial_y$ is the exterior derivative, commutes with $U(\bf{t})$:
\begin{align}
U(\bf{t})\left(\nabla \psi\right)=\nabla \left(U(\bf{t})\psi\right),
\end{align}
and because of this it also commutes with the Hamiltonian. It follows that $H$ is magnetic translation invariant. If we now consider the lattice $\mathbb{Z}^2$, we note that for $\bf{R}_1,\bf{R}_2\in \mathbb{Z}^2$, we have
\begin{align}
U(\bf{R}_1)U(\bf{R}_2) = U(\bf{R}_1+\bf{R}_2).
\end{align}
It follows that the operators $\{U(\bf{R}):\bf{R}\in\mathbb{Z}^2\}$ give a unitary representation of the lattice and $H$ is invariant under it. We can then apply Bloch's theorem, namely, look for eigenvectors of $H$ that are also simultaneous eigenvectors of the lattice magnetic translations. This will be wave functions $\psi_{\bf{k}}(\bf{r})$ satisfying
\begin{align}
U(\bf{R})\psi_{\bf{k}}(\bf{r})=e^{2\pi i \bf{k}\cdot \bf{R}}\psi_{\bf{k}}(\bf{r}),
\end{align}
which is equivalent to, setting $\bf{R}=(m,n)\in\mathbb{Z}^2$,
\begin{align}
\psi_{\bf{k}}(x+m,y+n)=e^{-2\pi i n x}e^{2\pi i\bf{k}\cdot \bf{R}}\psi_{\bf{k}}(x,y).
\label{eq: quasiperiodicity condition}
\end{align}
As usual, Bloch waves are not be normalizable in the whole of $\mathbb{R}^2$, but only in the unit cell of the lattice. The unit cell of the lattice, due to the above periodicity constraint should be understood, topologically, as a torus $\mathbb{R}^2/\mathbb{Z}^2$, and the wave functions at momentum $\bf{k}$, as sections of an appropriate line bundle over this torus.

We then want to look for wave functions in the lowest Landau level satisfying the quasiperiodicity condition in equation Eq.~\eqref{eq: quasiperiodicity condition}.
We now introduce the complex coordinate $w = x+iy$ in real space. We note that in this paper, the symbol $z_\tau = k_x + \tau k_y$ refers to momentum space complex coordinate, while $w$ refers to the real space complex coordinate. The covariant derivatives along the complex coordinate direction and its complex conjugate are
\begin{align}
    \nabla_w &= \frac{\partial}{\partial w} + \frac{\pi}{2}(w-\bar{w}) \quad \text{ and } \quad
    \nabla_{\bar{w}} = \frac{\partial}{\partial \bar{w}} + \frac{\pi}{2}(w-\bar{w}), 
\end{align}
and thus the Hamiltomnian looks
\begin{align}
H &=-\frac{1}{2}\left(\nabla_x\nabla_x+\nabla_y\nabla_y\right) =-\left(\nabla_{w}\nabla_{\bar{w}}+\nabla_{\bar{w}}\nabla_{w}\right)=-2\nabla_{w}\nabla_{\bar{w}}+\pi.
\end{align}
Noting that $-\nabla_w$ is the $L^2-$adjoint of $\nabla_{\bar{w}}$, the energy due to the first term for any state $|\psi\rangle$ is $\langle \psi|(-2\nabla_{w}\nabla_{\bar{w}})|\psi\rangle = 2||\nabla_{\bar{w}}\ket{\psi}||^2 \ge 0$ and so the ground state, namely the lowest Landau level, must satisfy $\nabla_{\bar{w}}\psi (\mathbf{r}) =0$, and the corresponding energy is $\pi$.

If we write $\psi(\mathbf{r}) = e^{-\pi y^2}\phi (\mathbf{r})$, the condition $\nabla_{\bar{w}}\psi (\mathbf{r}) = 0$ can be re-written as
\begin{align}
    \frac{\partial \phi (\mathbf{r})}{\partial \bar{w}} = 0, \label{eq: holomorphicity}
\end{align}
i.e. $\phi(\mathbf{r}) \equiv \phi (w)$ is holomorphic. The lowest Landau level is precisely the space of wave functions which are holomorphic in the sense of Eq.~\eqref{eq: holomorphicity} and square-integrable.

It turns out, as we shall prove below, that for fixed momentum $\bf{k}=(k_x,k_y)$, there is only one linearly independent solution consistent with the quasiperiodicity condition in equation Eq.~\eqref{eq: quasiperiodicity condition} given by 
\begin{align}
\psi_{\bf{k}}(\bf{r}) &=e^{-\pi y^2}e^{2\pi i k_x k_y} \thetachar{k_x}{-k_y}(w,i)
=
e^{-\pi y^2} \sum_{l\in\mathbb{Z}}e^{-\pi(l+k_x)^2 +2\pi i(l+k_x)w - 2\pi i lk_y}
\label{eq: LLL Bloch waves are theta functions}
\end{align}
One can explicitly check that the condition Eq.~\eqref{eq: quasiperiodicity condition} is satisfied by using the following identity for the theta function which holds for general $w_\tau = x + \tau y$ and $\gamma = m + \tau n$, where $\tau \in \mathbb{H}$ and $\mathbf{R} = (m,n)\in \mathbb{Z}^2$:
\begin{align}
\thetachar{a}{b}(w_{\tau}\!+\!\gamma,\tau) \!=\! e^{-i\pi\tau n^2 -2\pi i n w_{\tau}}e^{2\pi i (am -bn)}  \thetachar{a}{b}(w_{\tau},\tau).
\end{align}

The uniqueness of this solution (\ref{eq: LLL Bloch waves are theta functions}) at fixed momentum $\bf{k}$ can be seen in the following way.
As we have seen above, if we write $\psi_{\bf{k}}(\bf{r})=e^{-\pi y^2}\phi_\mathbf{k} (\bf{r})$, $\phi_\mathbf{k} (\bf{r})\equiv \phi_\mathbf{k} (w)$ is holomorphic in the complex variable $w$, and due to the condition (\ref{eq: quasiperiodicity condition}), it has to satisfy 
\begin{align}
\phi_\mathbf{k} (w+\gamma)=e_{\gamma}(w)e^{2\pi i(k_xm +k_yn)}\phi_\mathbf{k} (w), \label{eq:periodicityphi}
\end{align}
with the holomorphic multipliers
\begin{align}
e_{\gamma}(w):=e^{\pi n^2 -2\pi i nw},\ \gamma=m+ni \in\mathbb{Z}+i\mathbb{Z},    
\end{align}
Setting $\gamma=1$, we see that $\phi_\mathbf{k} (w+1)= \phi_\mathbf{k} (w)e^{2\pi i k_x}$. Hence we can expand it in Fourier series $\phi_\mathbf{k} (w)=\sum_{l\in\mathbb{Z}}c_l e^{2\pi i (l+k_x) w}$ and, working out the relations between the coefficients from Eq.~(\ref{eq:periodicityphi}), one obtains the recursive relation
\begin{align}
c_{l-n}=c_{l}e^{-\pi n^2 + 2\pi n (l+k_x) + 2\pi i k_y n}
\end{align}
whose general solution is $c_l=c e^{-\pi (l+k_x)^2 -2\pi i l k_y}$, where the prefactor $c$ may depend on $\mathbf{k}$. One then obtains the solution Eq.~\eqref{eq: LLL Bloch waves are theta functions}. We note that the holomorphic multipliers $e_\gamma (w)$ satisfy the co-cycle condition $e_{\gamma_1+\gamma_2}(w)=e_{\gamma_2}(w+\gamma_1)e_{\gamma_1}(w)$. One can see the uniqueness of the solution (\ref{eq: LLL Bloch waves are theta functions}) also from the fact that such holomorphic multipliers define a holomorphic line bundle with degree $1$ (1st Chern number) over the unit cell, which has a unique holomorphic section by the Riemann-Roch theorem.

Finally we look for the expression of the cell (quasi)-periodic part of the Bloch wave, denoted by $u_\mathbf{k} (\mathbf{r})$, which is defined by
\begin{align}
\psi_{\bf{k}}(\bf{r})= e^{2\pi i\bf{k}\cdot \bf{r}}u_{\bf{k}}(\bf{r}).
\end{align}
The plane wave part of the above formula makes it so that the quasi-periodicity condition from Eq.(\ref{eq: quasiperiodicity condition}) that $u_{\bf{k}}(\bf{r})$ must satisfy has the coefficient independent of $\mathbf{k}$:
\begin{align}
u_{\bf{k}}(x+m,y+n)=e^{-2\pi i n x}u_{\bf{k}}(x,y),\ m,n\in\mathbb{Z},
\label{eq: quasiperiodicity periodic part}
\end{align}
so they can all be interpreted as (square integrable) sections of the same line bundle over the unit cell torus, hence they belong to the same Hilbert space---see also the main text on the definition of a Bloch wave function, which uses $\ket{u_{\bf{k}}}\in\mathcal{H}$ for fixed $\mathcal{H}$ independently of $\bf{k}$. From the explicit expression of $\psi_\mathbf{k}(\mathbf{r})$ in Eq.~(\ref{eq: LLL Bloch waves are theta functions}), we see that
\begin{align}
    u_{\bf{k}}(x,y) = e^{-2\pi i\mathbf{k}\cdot\mathbf{r}} e^{-\pi y^2}e^{2\pi i k_x k_y} \thetachar{k_x}{-k_y}(w,i).
\end{align}
This expression of the Bloch wave function of the lowest Landau level should coincide with the general expression obtained in Eq.~\eqref{eq: LLL Bloch wave functions} in Sec.~\ref{sec: technical results} with the replacement $\mathcal{C} = 1$ and $\tau = i$. So far, they do not look the same; the characteristics of the theta function is the momentum for the former and real-space coordinate for the latter. They are indeed the same and we are going to show it by interchanging the characteristics and the argument of the theta function. We use the notation $z = k_x + ik_y$ from the previous section to describe the complex coordinate in momentum space.

We first note that the following identity for theta functions is known:
\begin{align}
\thetachar{a}{b}(w_\tau,\tau)=e^{i\pi\tau a^2} e^{2\pi i a(w_{\tau}+b)}\thetachar{0}{0}(w_{\tau}+b+\tau a,\tau).
\end{align}
This identity, together with Jacobi identity of the basic theta function under modular transformations, $\thetachar{0}{0}(w_{\tau}/\tau,\tau) = (-i\tau)^{\frac{1}{2}} e^{\frac{\pi}{\tau} i w_\tau^2}\thetachar{0}{0}(w_{\tau},\tau)$, we can interchange the characteristics and the argument and rewrite the expression in the following way:
\begin{align}
    u_{\bf{k}}(x,y) 
    &=
    e^{-2\pi i\mathbf{k}\cdot\mathbf{r}} e^{-\pi y^2}e^{2\pi i k_x k_y} e^{-\pi k_x^2 + 2\pi i k_x (w - k_y)} \thetachar{0}{0}(w+iz,i)
    \notag \\
    &=
    e^{-2\pi i\mathbf{k}\cdot\mathbf{r}} e^{-\pi y^2}e^{2\pi i k_x k_y} e^{-\pi k_x^2 + 2\pi i k_x (w - k_y)}e^{-\pi (w+iz)^2} \thetachar{0}{0}(z-iw,i)
    \notag \\
    &=
    e^{-2\pi i\mathbf{k}\cdot\mathbf{r}} e^{-\pi y^2}e^{2\pi i k_x k_y} e^{-\pi k_x^2 + 2\pi i k_x (w - k_y)}e^{-\pi (w+iz)^2}e^{\pi x^2}e^{2\pi i x (z+y)} \thetachar{-x}{y}(z,i)
    \notag \\
    &= e^{2\pi i k_x k_y - \pi k_y^2} \thetachar{-x}{y}(z,i).
\end{align}
This is nothing but the expression obtained in Eq.~\eqref{eq: LLL Bloch wave functions} in Sec.~\ref{sec: technical results} for $\mathcal{C} = 1$ and $\tau = i$. This completes our proof that the lowest Landau level wave function is precisely the unique geometrically flat K{\"a}hler band with $\mathcal{C} = 1$ and $\tau = i$.
\end{widetext}
\end{document}